\documentclass[3p,11pt,preprint]{elsarticle}

\pdfoutput=1

\usepackage{natbib}
\biboptions{numbers,square,comma,sort&compress}
\bibpunct{[}{]}{,}{n}{,}{,}

\usepackage{amsmath,amssymb}
\usepackage{url}
\usepackage{algorithm}
\usepackage[noend]{algpseudocode}
\usepackage[table]{xcolor}
\usepackage{tikz}
\usetikzlibrary{arrows}
\usepackage{multirow}

\newcolumntype{C}[1]{>{\centering\let\newline\\\arraybackslash\hspace{0pt}}m{#1}}

\journal{Nucl.~Instr.~Meth.~Phys.~Res.~A}

\begin{document}

\begin{frontmatter}

\title{Multi-objective shape optimization of radio frequency cavities using an evolutionary algorithm}

\address[ETH]{Computer Science Department, ETH Zurich, 8092 Zurich, Switzerland}
\address[PSI]{Paul Scherrer Institute (PSI), 5232 Villigen, Switzerland}

\author[ETH]{Marija Kranj\v{c}evi\'c\corref{mycorrespondingauthor}}
\cortext[mycorrespondingauthor]{Corresponding author}
\ead{marija.kranjcevic@inf.ethz.ch}

\author[PSI]{Andreas Adelmann}
\ead{andreas.adelmann@psi.ch}
\author[ETH]{Peter Arbenz}
\ead{arbenz@inf.ethz.ch}
\author[PSI]{Alessandro Citterio}
\ead{alessandro.citterio@psi.ch}
\author[PSI]{Lukas Stingelin}
\ead{lukas.stingelin@psi.ch}

\begin{abstract}
Radio frequency (RF) cavities are commonly used to accelerate charged particle beams. The shape of the RF cavity determines the resonant electromagnetic fields and frequencies, which need to satisfy a variety of requirements for a stable and efficient acceleration of the beam. For example, the accelerating frequency has to match a given target frequency, the shunt impedance usually has to be maximized, and the interaction of higher order modes with the beam minimized. In this paper we formulate such problems as constrained multi-objective shape optimization problems, use a massively parallel implementation of an evolutionary algorithm to find an approximation of the Pareto front, and employ a penalty method to deal with the constraint on the accelerating frequency. Considering vacuated axisymmetric RF cavities, we parameterize and mesh their cross section and then solve time-harmonic Maxwell's equations with perfectly electrically conducting boundary conditions using a fast 2D Maxwell eigensolver. The specific problem we focus on is the hypothetical problem of optimizing the shape of the main RF cavity of the planned upgrade of the Swiss Synchrotron Light Source (SLS), called SLS-2. We consider different objectives and geometry types and show the obtained results, i.e., the computed Pareto front approximations and the RF cavity shapes with desired properties. Finally, we compare these newfound cavity shapes with the current cavity of SLS.
\end{abstract}


\end{frontmatter}


\section{Introduction}
Radio frequency (RF) cavities have been used in particle accelerators since Ising proposed resonant acceleration in 1924~\cite{ising1924prinzip}, Wider\"oe  built the first linear accelerator (LINAC) in 1928~\cite{Wideroe:1928flt}, and Lawrence and Livingston built the first circular accelerator, the cyclotron, in 1930~\cite{PhysRev.40.19}. They are used to accelerate charged particles, from electrons and protons to heavy ions and uranium~\cite{Yamada:2008zzb}. At the Paul Scherrer Institute~(PSI), for example, RF cavities with resonant frequencies of 50 and 150~MHz are used to accelerate a proton beam of 2.2~mA up to an energy of 590~MeV to drive a spallation neutron source and a muon and pion source for fundamental research. Four coupled cavities with resonant frequencies of 70~MHz power the cyclotron of the PROSCAN facility for cancer treatment.  In the storage ring of the Swiss Synchrotron Light Source (SLS), four 500~MHz ELETTRA-type~\cite{Fernandes:1989dn} cavities compensate for the energy loss due to synchrotron radiation and focus the electron bunches in longitudinal direction. A pair of passive superconducting cavities  with operation frequency at the third harmonic provide bunch lengthening and Landau damping~\cite{Svandrlik:2000gf}. The newest accelerator at PSI is the SwissFEL facility~\cite{PhysRevAccelBeams.19.100702}, with pulsed RF cavities  at operation frequencies of 3, 5.7 and 12~GHz. Compared to other PSI facilities, where standing wave cavities are used, most of the cavities of the SwissFEL LINAC are of the traveling wave cavity type. Other than for beam acceleration, RF cavities are also used for beam diagnostics, like beam current and position monitors~\cite{Sun:2015eta, Keil:2016pqq}, for bunch rotation in deflecting cavities~\cite{Marchetti:2017eym}, or in power amplifiers of accelerators, like klystrons, or filters.

PSI is currently elaborating an upgrade proposal for SLS, called SLS-2~\cite{SLS2CDR2017}, for dramatically improved synchrotron light quality. This process is triggered by the development of distributed vacuum pumping, which allowed a reduction of the vacuum chamber dimensions, larger number and more compact magnets, and advancements in high precision machining. It is foreseen to reuse the existing main cavities, extended by a new type of absorber for the most troubling beam excited higher order modes (HOMs).

In order to efficiently accelerate the beam, the RF cavity has to be optimized and engineered to satisfy a variety of requirements. Accelerator physics requirements like longitudinal focusing, space limitations and the availability of power sources usually define the desired resonant frequency, the voltage and the number of cavities. The shunt impedance~\cite{wiedemann2015particle} has to be maximized in order to reduce the power load. To avoid discharges and excessive dark current in the RF cavity, the electric field on the cavity surface should not exceed a critical threshold, and, to avoid thermal gradients and local heating, maximum current densities should not be exceeded. Furthermore, the interaction of HOMs with the beam should be minimized, and multipacting on the surfaces avoided. In high-intensity LINACs and synchrotrons it is also desirable to increase the quality factor in order to reduce beam loading effects. In order to enable manufacturing, overly complicated shapes and tight tolerances should be avoided. Additionally, tuning the cavity resonant frequency and coupling with the amplifier usually 
must be possible.

Since the resonant modes, electromagnetic (EM) fields and frequencies are determined by the shape of the RF cavity, the goal is to find a cavity shape, or shapes, that satisfy the above requirements. This can be formulated as a multi-objective shape optimization problem where, for example, an objective function could be defined as the peak value of the electric field on the cavity surface, and the corresponding objective to have this value as small as possible.

The design of the first RF cavities was done by analytically solving the eigenvalue problem arising from Maxwell's equations, perturbation corrections~\cite{jackson_classical_1999} and mode matching techniques. Modern design of RF cavities is done with the help of computer programs, like SUPERFISH~\cite{Halbach:1976cp} for axisymmetric cavities, FemaXX~\cite{GeusPhd} or commercially available codes like HFSS~\cite{hfss} and Microwave Studio~\cite{MicrowaveStudio} for complex three-dimensional cavities, and coupled multiphysics codes like ANSYS~\cite{ansys} or COMSOL~\cite{comsol} for simulating the cavity deformation due to heat load and air pressure~\cite{InjectorIIResonatorCC07}. 

Different approximation methods, such as the finite difference method, the boundary element method or the finite element method (FEM) can be used to solve time-harmonic Maxwell's equations in order to compute the necessary EM fields and frequencies. Due to its flexibility for modeling the problem in terms of domain approximation and favorable previous results, both for the three-dimensional~\cite{GeusPhd} and the axisymmetric case~\cite{ChinellatoPhd}, in this paper we opt for the mixed FEM, and solve the resulting generalized eigenvalue problem using the symmetric Jacobi--Davidson algorithm~\cite{arbenz:01,GeusPhd}. 

The optimization of the cavity shape is still usually done by starting from strongly simplified geometries which possess analytical solutions and symmetries. With the help of some intuition from perturbation theory, the geometry is then iteratively optimized until a `good enough' geometry is found. Then, for example, the objective functions are scalarized, i.e., the mul\-ti-ob\-jec\-tive optimization problem is converted into a single-objective optimization problem, usually with the help of some predetermined weights, and a gradient-based optimization method is applied, starting from the already found cavity shape~\cite{Schaer:2016jhu}. Approaches that employ scalarization and gradient-based methods for EM shape optimization problems where the frequency of the accelerating mode has to match a given target frequency and one or a few properties of the EM field of the three-dimensional cavity shape need to be optimized are published in~\cite{akcelik:05,akcelik:08,akcelik:09}.

Another way of dealing with multiple objectives is to optimize them separately, with algorithms such as simulated annealing~\cite{kirkpatric:83}, particle swarm optimization~\cite{kennedy:95}, or evolutionary algorithms (EAs)~\cite{kalyanmoy-deb,dan-simon}. EAs are probably the most popular, and they have already been successfully applied in the area of particle accelerator physics. For example, in~\cite{bazarov:05} an EA was used to optimize a high brightness dc photoinjector, and~\cite{ineichen:13} presented a massively parallel multi-objective optimization tool combined with beam dynamics simulation, which employed an EA as the optimization method. Because of conflicting objectives, the ability of EAs to escape local optima and deal with possibly discontinuous objective functions, as well as their suitability for parallelization, we use an EA to find a set of candidate solutions that represent potentially interesting cavity shapes.

In this paper, in order to illustrate the use of an EA for RF cavity shape optimization, we focus on the hypothetical problem of optimizing the shape of the main RF cavity of SLS-2. To simplify the problem and reduce the computation time, we consider axisymmetric cavities and a set of four objectives: minimizing the error between the frequency of the accelerating mode and a given target frequency, maximizing the shunt impedance, minimizing the peak electric field on the cavity surface and avoiding beam interaction with HOMs. Note that the first objective can also be formulated as the constraint that the frequency of the accelerating mode has to match the target frequency. While the first three objective functions possess a certain level of smoothness~\cite{RevModPhys.18.441}, the beam interaction objective function in its most natural formulation is discontinuous which supports the use of an EA. Cavities with coaxial coupling~\cite{Schaer:2016jhu} or antennas with neglectable field perturbation intrinsically satisfy the axisymmetry condition. In cases where the axisymmetry condition is broken by, for example, a coupler or a damper, a fully three-dimensional solver needs to be used, but, at the expense of increased computational cost, the same optimization method could still be employed.

To formulate a multi-objective optimization problem we need to parameterize the cavity shape by a finite number of design variables. Since axisymmetric shapes are determined by their cross section, it is enough to parameterize the latter. Standard approaches for parameterizing admissible shapes include defining the design variables as positions of some boundary nodes of a mesh~\cite{francavilla,le} or as parameters that define the curve which specifies the boundary of the shape, such as, for example, polynomials~\cite{bhavikatti}, cubic splines~\cite{tortorelli-2}, or B-splines~\cite{braibant}. Since the first approach results in many design variables and often needs additional constraints to ensure the regularity of the boundary~\cite{haslinger,braibant}, we opt for the second one, focusing in particular on curves which define shapes that can be manufactured and are known to give good results, such as ellipse arcs, which generate cavity shapes similar to the ELETTRA-type cavity. Additionally, we consider boundaries defined by complete cubic splines in order to explore a wider variety of potentially interesting cavity shapes.

This paper is organized as follows. In section~\ref{sec:multi-objective-optimization} we describe the parameterization of the cavity cross section and give a formal definition of the multi-objective shape optimization problem. Section~\ref{sec:forward-solver} deals with the solution of time-harmonic Maxwell's equations and the computation of the objective function values. Section~\ref{sec:EA} describes the EA, as well as our treatment of constraints. In section~\ref{sec:results} we solve several multi-objective shape optimization problems for SLS-2, considering two different geometry types. Finally, in section~\ref{sec:conclusions}, we draw conclusions and discuss possible future work. 


\section{Multi-objective optimization}\label{sec:multi-objective-optimization}


\subsection{Parameterizing the geometry}\label{sec:parameterization}

\begin{figure}[h!]
\includegraphics[width=\textwidth,trim={0cm 5.1cm 0cm 3.9cm}]{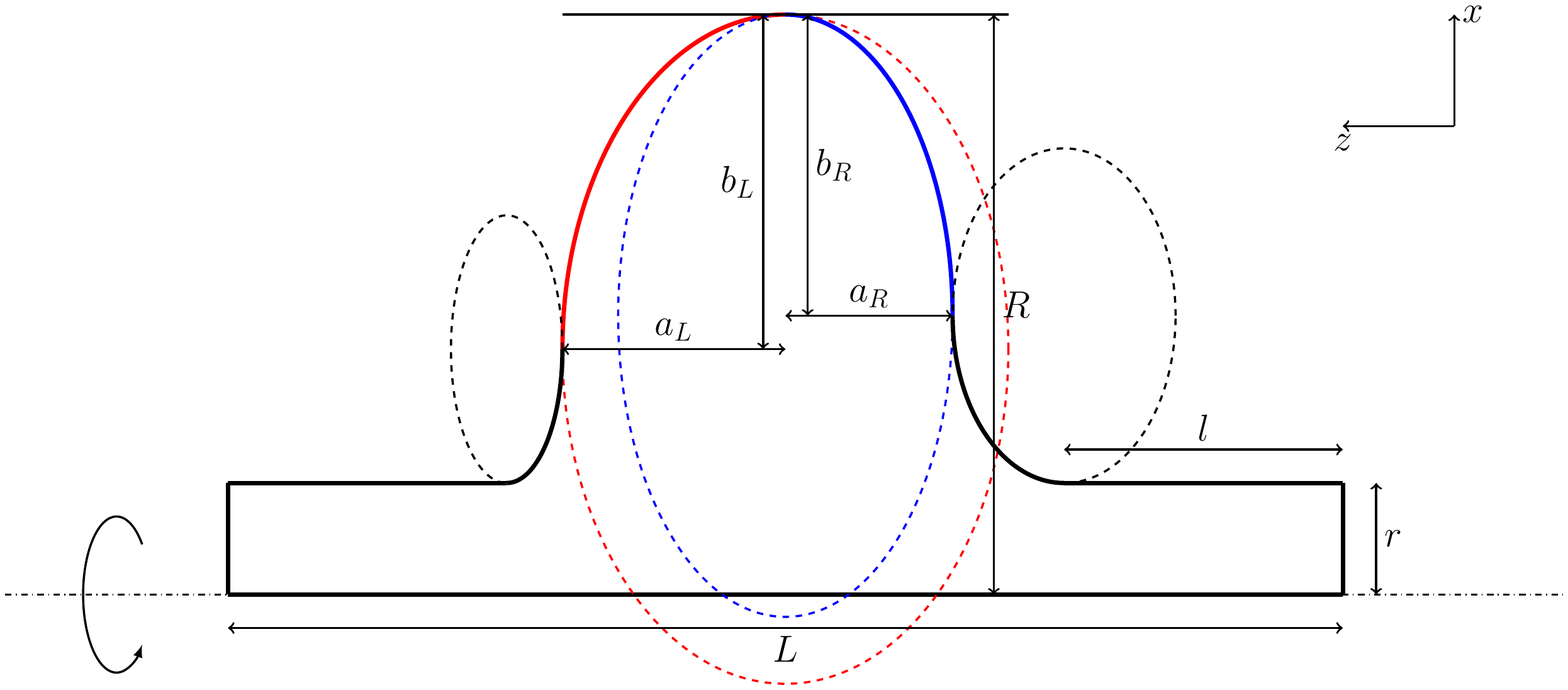}
\caption{An axisymmetric geometry type, with the cross section parameterized by $L$, $l$, $R$, $r$, $a_L$, $b_L$, $a_R$, and $b_R$. The curved part of the boundary consists of four ellipse arcs of 90$^\circ$.}
\label{fig:parameterization}
\end{figure}

We consider fixed axisymmetric geometry types that can be parameterized by some design variables $d_1,\dots,d_N$. An example of a geometry type with an asymmetric cross section, parameterized by $N=8$ variables, is shown in figure~\ref{fig:parameterization}. The variables $L$, $l$ and $r$ specify the lengths of the straight parts of the cross section boundary. The curved part of the boundary comprises four ellipse arcs of $90^\circ$. The semi axes of the upper left and right ellipse are given by $a_L$, $b_L$, and $a_R$, $b_R$, respectively, and their highest point is determined by $R$.


\subsection{Multi-objective optimization problem}\label{sec:MOOP}
Having fixed a parameterizable geometry type, we consider the multi-objective optimization problem
\begin{align}
& \min F_i(\boldsymbol{d}),\hspace{8pt} i=1,\dots,n,\nonumber\\
& G_i(\boldsymbol{d}) \geq 0,\hspace{8pt} i = 1,\dots,k,\label{eq:G}\\
& H_i(\boldsymbol{d}) = 0,\hspace{8pt} i = 1,\dots,l,\label{eq:H}
\end{align}
where $\boldsymbol{d} = (d_1,\dots,d_N)^T$ $\in X \subset \mathbb{R}^N$ is a design point, $F_1,\dots,F_n : X \to \mathbb{R}$ are objective functions, and (\ref{eq:G}) and (\ref{eq:H}) are inequality and equality constraints, respectively, with $$G_1,\dots,G_k,H_1,\dots,H_l : X \to \mathbb{R}.$$
For example, the objective functions could be defined as properties of the the accelerating mode, such as the peak electric field on the cavity surface, or, since the shunt impedance $R_{sh}^{(0)}$ should be maximized, $-R_{sh}^{(0)}$.
A possible equality constraint could be that the frequency of the accelerating mode $f^{(0)}$ should match a given target frequency $f_{RF}$, i.e.,
\begin{align}
H(\boldsymbol{d}) = \big|f^{(0)}(\boldsymbol{d}) - f_{RF}\big| = 0,
\label{eq:frequency-constraint}
\end{align}
where $f^{(0)}(\boldsymbol{d})$ indicates that $f^{(0)}$ depends on the cavity shape determined by $\boldsymbol{d}$.


\tikzstyle{b} = [rectangle, draw, fill=gray!30!white, text width=8em,
                 text centered, rounded corners, minimum height=2em, thick]
\tikzstyle{l} = [draw, -latex', thick]

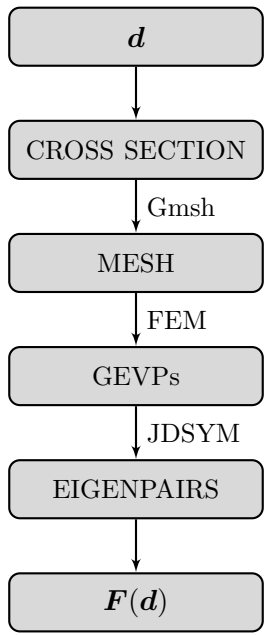
\begin{figure}[b!]
\begin{center}
\begin{tikzpicture}[scale=1]
\node [b] (node-d) {$\boldsymbol{d}$};
\node [b] (cross-section) at ([shift={(0,-1.5)}] node-d) {\small CROSS SECTION};     
\path [l] (node-d)  -- (cross-section);
\node [b] (mesh) at ([shift={(0,-3)}] node-d) {\small MESH};  
\path [l] (cross-section)  -- (mesh) node[midway,right] {\small Gmsh};
\node [b] (gevps) at ([shift={(0,-4.5)}] node-d) {\small GEVPs};
\path [l] (mesh)  -- (gevps) node[midway,right] {\small FEM};
\node [b] (eigenpairs) at ([shift={(0,-6)}] node-d) {\small EIGENPAIRS};
\path [l] (gevps)  -- (eigenpairs) node[midway,right] {\small JDSYM};
\node [b] (node-f) at ([shift={(0,-7.5)}] node-d) {$\boldsymbol{F}(\boldsymbol{d})$};
\path [l] (eigenpairs)  -- (node-f);
\end{tikzpicture}
\end{center}\vspace{-10pt}
\caption{The basic steps for evaluating the vector objective function $\boldsymbol{F} = (F_1,\dots,F_n)^T$ in a given design point $\boldsymbol{d}$.}
\label{fig:forward-solver}
\end{figure}

\section{Forward solver}\label{sec:forward-solver}
\begin{figure}[h!]
\includegraphics[width=\textwidth,trim={0cm 5.2cm 0cm 3.8cm}]{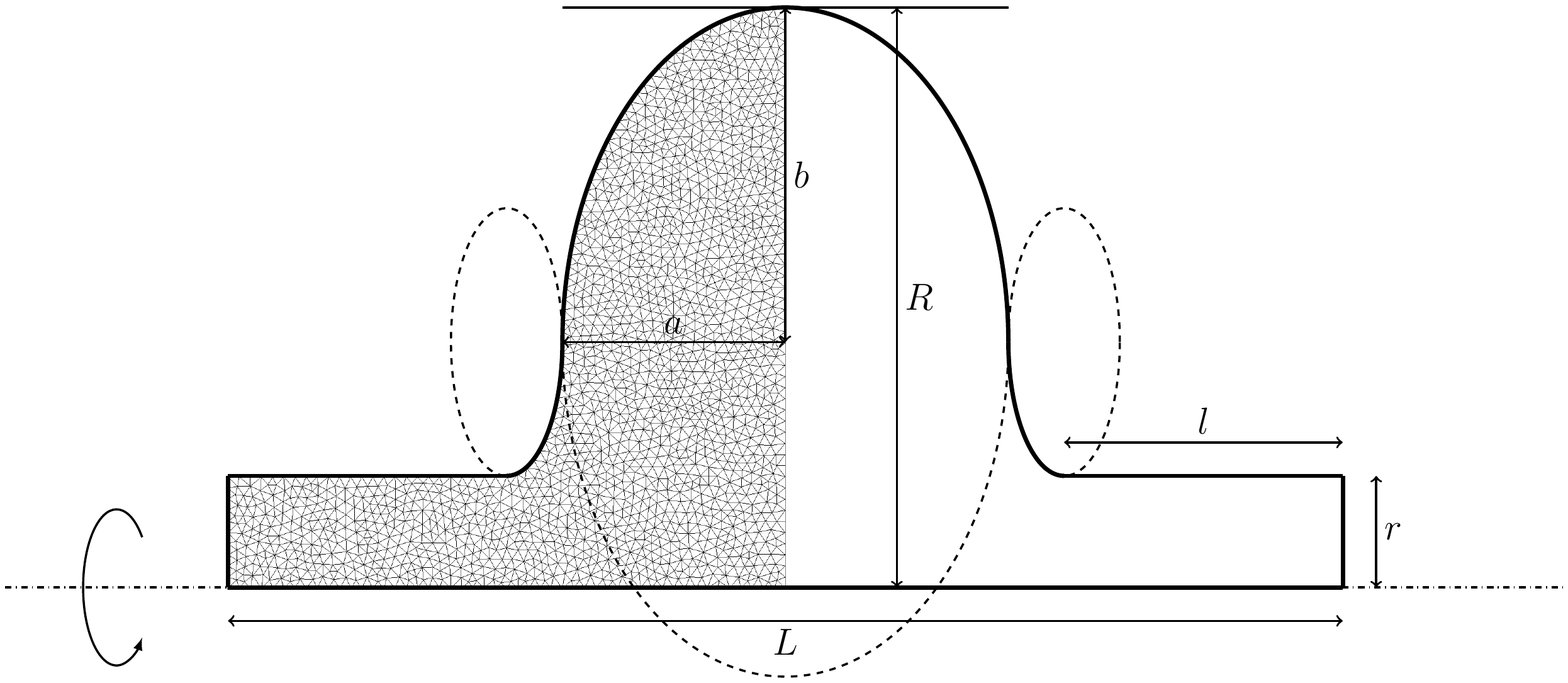}
\caption{A geometry type with a symmetric cross section, defined by $L$, $l$, $R$, $r$, $a$, and $b$. We consider $L$, $l$, and $r$ to be fixed, and $R$, $a$, and $b$ to be design variables, i.e., $\boldsymbol{d}=(R,a,b)^T$. Since this cross section is symmetric, it is enough to mesh half of it.}
\label{fig:sym-elliptical-mesh}
\end{figure}

A way to evaluate the objective functions in a given design point, i.e., to compute $$\boldsymbol{F}(\boldsymbol{d}) = \big(F_1(\boldsymbol{d}),\dots,F_n(\boldsymbol{d})\big)^T,$$ is illustrated in figure~\ref{fig:forward-solver}.
As shown in section~\ref{sec:parameterization}, once the type of the geometry is fixed, a design point determines the shape of the cavity by determining its cross section. We compute the resonant modes of this cavity by solving time-harmonic Maxwell's equations, assuming that there is vacuum inside the cavity and that the boundary conditions are perfectly electrically conducting (PEC).\\
We use the Gmsh~\cite{gmsh} C++ API to create unstructured triangular meshes of the cross sections.
In addition to the geometry type shown in figure~\ref{fig:parameterization}, a wide range of geometry types can easily be created using lines, ellipses and, for example, non-uniform rational B-splines.
Using the mixed FEM, for each $m\in\mathbb{N}_0$ we get a generalized eigenvalue problem (GEVP) of the form~\cite[p.70]{ChinellatoPhd}
\begin{align}
\boldsymbol{A}^{(m)}\hspace{1pt}\boldsymbol{q} = \lambda\hspace{1pt}\boldsymbol{M}^{(m)}\hspace{1pt} \boldsymbol{q}, \hspace{8pt}
\Big(\boldsymbol{Y}^{(m)}\Big)^T\boldsymbol{M}^{(m)}\hspace{1pt}\boldsymbol{q} = 0,
\label{eq:GEVP}
\end{align}
where $\boldsymbol{A}^{(m)}$ is symmetric positive semidefinite, $\boldsymbol{M}^{(m)}$ is symmetric positive definite, and $\boldsymbol{Y}^{(m)}$ is the null space of $\boldsymbol{A}^{(m)}$. However, in practice, when computing a few of the lowest resonant modes, only a few of these GEVPs need to be solved and, for each eigenproblem, only the eigenpairs corresponding to the few smallest non-zero eigenvalues have to be found. We solve these GEVPs with the symmetric Jacobi--Davidson (JDSYM) algorithm~\cite{jdsym-trilinos,abgh:04} implemented using Trilinos~\cite{trilinos}. From the computed eigenpairs we easily calculate the required properties of the corresponding EM field.
If the cavity cross section is symmetric (see figure~\ref{fig:sym-elliptical-mesh}), it is enough to solve time-harmonic Maxwell's equations in only half of the cross section, once with PEC and once with perfectly magnetically conducting (PMC) conditions on the symmetry plane~\cite{jin:93}.
The values of the EM fields in the entire cross section can then easily be obtained from the computed results using symmetry.

To give an impression on the computation work, on one core of Intel XeonE5\_2680v3, when computing a few of the lowest modes of an axisymmetric approximation of the ELETTRA-type cavity 
using half of its cross section, it takes 12.4 seconds to construct a mesh of the cross section which comprises 158'308 triangles, 28.2 seconds to create the matrices from (\ref{eq:GEVP}) (for $m=0$) and find the eigenpairs corresponding to the three smallest non-zero eigenvalues (with the tolerance $10^{-10}$), and 1.2 seconds to compute properties of the EM field, including the peak electric field on the cavity surface, the shunt impedance and the quality factor.


\section{Evolutionary algorithm (EA)}\label{sec:EA}
Since the minimizers of different objectives are usually different points,
we use the concept of dominance to be able to compare candidate solutions~\cite[p.519]{dan-simon}. A point $\boldsymbol{d}_1$ dominates another point $\boldsymbol{d}_2$ if it is not worse than $\boldsymbol{d}_2$ in any of the objectives, and it is strictly better in at least one objective. We search for points that are not dominated by any other point, called Pareto optimal points, using a massively parallel implementation of an EA from~\cite{ineichen:13}. 
The main steps of an EA are shown in algorithm~\ref{alg:EA}. The population of the first generation comprises $M$ individuals, each possessing design variables, called genes in the context of an EA, with randomly chosen values (line~\ref{alg:EA-initialize}).
In line~\ref{alg:EA-evaluate} the objective functions are evaluated in the design points corresponding to the individuals in the population.
The algorithm then performs a predetermined number of cycles, each resulting in a new generation (lines~\ref{alg:EA-cycle}-\ref{alg:EA-selector2}). In every cycle, new individuals are created using two operators: crossover, which models the exchange of genetic material between homologous chromosomes, and mutation, which models accidental changes in the set of genes (lines~\ref{alg:EA-crossover1},\ref{alg:EA-crossover2}). The new individuals are evaluated (line~\ref{alg:EA-evaluate-new}) and, out of these new individuals and the individuals in the current generation, approximately $M$ fittest ones are chosen to form a new generation (lines~\ref{alg:EA-selector1},\ref{alg:EA-selector2}). In case of choosing between two equally fit individuals, a crowding distance measure is used to prevent niches. We use PISA NSGA-II~\cite{nsga-ii,pisa} for this selection process.

\begin{algorithm}
\caption{Evolutionary algorithm}\label{alg:EA}
\begin{algorithmic}[1]
\State initialize a random population of individuals, $I_i$, $i = 1, \dots, M$ \label{alg:EA-initialize}
\State evaluate the population \label{alg:EA-evaluate}
\For {a predetermined number of generations} \label{alg:EA-cycle}
\State \textcolor{gray}{// \emph{variator}}
\For {pairs of individuals $I_i$, $I_{i+1}$} \label{alg:EA-crossover1}
\State $crossover(I_i, I_{i+1})$, and, with some probability, $mutation(I_i)$, $mutation(I_{i+1})$ \label{alg:EA-crossover2}
\EndFor
\State evaluate new individuals \label{alg:EA-evaluate-new}
\State \textcolor{gray}{// \emph{selector}}
\State sort individuals according to levels of dominance \label{alg:EA-selector1}
\State choose $M$ fittest individuals to form the next generation \label{alg:EA-selector2}
\EndFor
\end{algorithmic}
\end{algorithm}


\subsection{Frequency constraint}\label{sec:frequency-constraint}
Since it is virtually impossible that the target frequency $f_{RF}$ will be matched exactly when a numerical method is used for computing the solution, we replace the constraint (\ref{eq:frequency-constraint}) with
\begin{align}
H(\boldsymbol{d}) = \big|f^{(0)}(\boldsymbol{d}) - f_{RF}\big| \leq \varepsilon,
\label{eq:soft-frequency-constraint}
\end{align}
for some small positive constant $\varepsilon$. We then employ a non-death-penalty approach~\cite[p.486]{dan-simon}, i.e., we keep the individuals which violate the constraint (\ref{eq:soft-frequency-constraint}) in the population, but penalize their objective function values (algorithm~\ref{alg:frequency-constraint}). The values of the penalty tolerance $\varepsilon$ and the penalty factor $\alpha$ have to be chosen appropriately.
\begin{algorithm}
\caption{Penalization}\label{alg:frequency-constraint}
\begin{algorithmic}[1]
\If {$\big|f^{(0)}(\boldsymbol{d})-f_{RF}\big| > \varepsilon$} \label{alg:constraint-violated}
\For {$i = 1,\dots, n$}
\State $ F_i(\boldsymbol{d}) \mapsto F_i(\boldsymbol{d}) + \alpha \cdot \big|f^{(0)}(\boldsymbol{d})-f_{RF}\big|$
\EndFor
\EndIf
\end{algorithmic}
\end{algorithm}


\section{Results}\label{sec:results}
\begin{figure}[h!]
\includegraphics[width=\textwidth,trim={1.5cm 8cm 1.cm 4.0cm}]{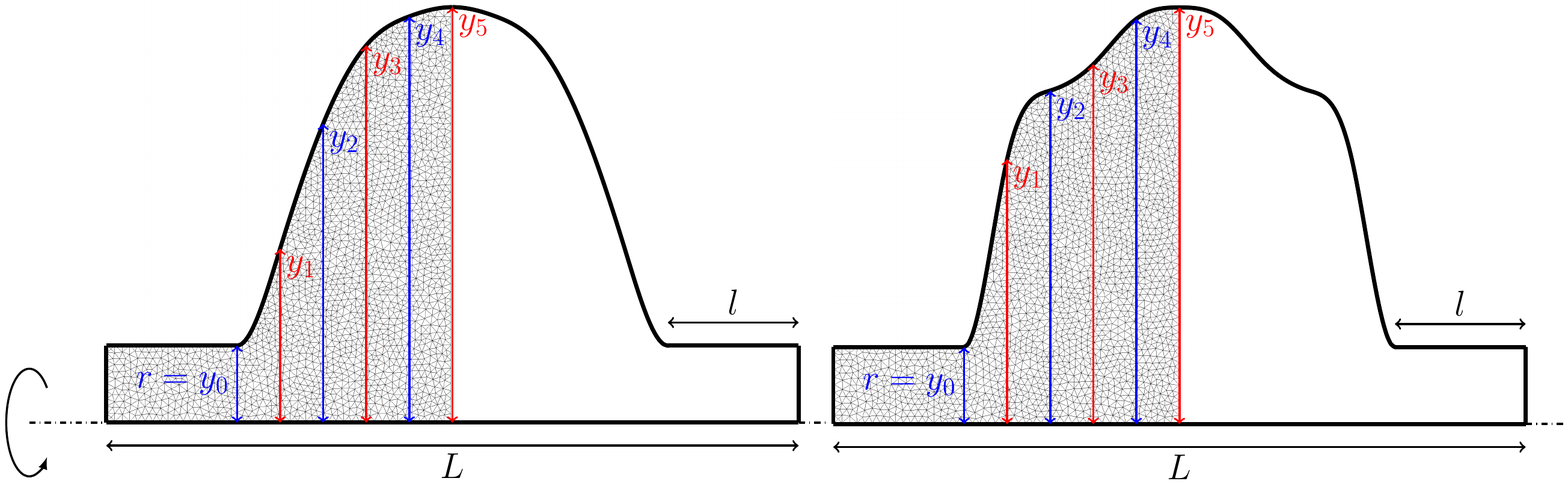}
\caption{A geometry type with a symmetric cross section whose boundary is a complete cubic spline with $n$ equidistant knots and horizontal end slopes. A geometry is defined by $L$, $l$, $r=y_0$, and $y_1,\dots,y_{n-1}$. We set $n=6$, fix the values of $L$, $l$, and $r$ and take $y_1,\dots,y_5$ to be design variables, i.e., $\boldsymbol{d}=(y_1,\dots,y_5)^T$. Different values of design variables result in different instances of the same parameterization: the one on the left resembles the elliptical cavity from figure~\ref{fig:sym-elliptical-mesh}, while the one on the right has a more complicated shape. As in figure~\ref{fig:sym-elliptical-mesh}, only half of the cross section needs to be meshed.}
\label{fig:splines}
\end{figure}

We focus on the hypothetical optimization of the shape of the RF cavity for SLS-2.
We consider different objective functions and two different geometry types with a symmetric cross section: an elliptical geometry type, illustrated in figure~\ref{fig:sym-elliptical-mesh}, and the geometry type where half of the cross section is defined as a complete cubic spline with horizontal end slopes, shown in figure~\ref{fig:splines}. 
In both of these cases, in order to satisfy accelerator requirements we fix the value $r = 50$~mm, and, to have enough space for the modes to decay before the end of the beam pipe we set $l = 188.671$~mm and $L = 680$~mm.
We use the blend crossover and the independent bit mutation~\cite[p.27]{IneichenPhd}, the initial values of design variables are chosen randomly from a given interval, and the number of individuals in a generation is always around $M = 100$.
We run all the optimizations on the Euler cluster of ETH~Zurich, with gcc 4.8.2, Gmsh 2.12.0, and Trilinos 12.6.1.
The description of interesting individuals is given in table~\ref{table:individuals-description}. We check the stability of the results by further mesh refinement and show the computed objective function values in table~\ref{table:individuals-comparison}. 
We compare these values with the corresponding values of an axisymmetric approximation of the current cavity of SLS, denoted ELETTRA~2D in table~\ref{table:individuals-comparison}. 

We first consider the elliptical geometry type from figure~\ref{fig:sym-elliptical-mesh}. Having fixed the variables $L$, $l$, and $r$, this geometry type is defined by three design variables: $R$, the equator radius, and $a$ and $b$, the ellipse half axes. In the following examples the initial values of these design variables are (in mm) $R\in[200,350]$, and $a,b\in[50,150]$.

\renewcommand{\arraystretch}{1.25}
\begin{table}[b!]
\centering \small
\caption{A description of chosen individuals.}\vspace{2pt}
\label{table:individuals-description}
\begin{tabular}{|c|c|C{1.5cm}|C{1.92cm}|c|}
\hline
NAME & GEOMETRY TYPE & FROM & SHOWN IN & DESIGN VARIABLES \\ \hline\hline
CAVITY~\#1 & figure~\ref{fig:sym-elliptical-mesh} & figure~\ref{fig:geom-type-1-gen} & figure~\ref{fig:geom-type-1-id} & $R = 249.901$, $a = 125.232$, $b = 70.2322$ \\ \hline
CAVITY~\#2 & figure~\ref{fig:sym-elliptical-mesh} & figure~\ref{fig:geom-type-1-gen-4obj} & figure~\ref{fig:geom-type-1-id-4obj} & $R = 251.972$, $a = 121.887$, $b = 78.5213$ \\ \hline
\multirow{2}{*}{CAVITY~\#3} & \multirow{2}{*}{figure~\ref{fig:splines}} & \multirow{2}{*}{-} & \multirow{2}{*}{figure~\ref{fig:geom-type-3-id2-4obj}} & $y_1 = 141.759$, $y_2 = 227.387$, $y_3 = 246.357$ \\ 
                  &                   &                   &                   & $y_4 = 254.171$, $y_5 = 257.5$ \\ \hline
\end{tabular}
\end{table}
\begin{table}[b!]
\centering \small
\caption{A comparison of individuals described in table~\ref{table:individuals-description}. The values in shaded fields were optimized for, and the other values are shown for a more thorough comparison.}\vspace{2pt}
\label{table:individuals-comparison}
\begin{tabular}{|c|c|c|c|c|c|c|}
\hline
NAME & $f^{(0)}\hspace{1pt}\big[$MHz$\big]$ & $F_3\hspace{1pt}\big[$1$/$m$\big]$ & $R_{sh}^{(0)}\hspace{1pt}\big[$M$\Omega\big]$ & $R_{sh}^{(0)}/Q_{0}^{(0)}\hspace{1pt}\big[\Omega\big]$ & LIST OF DANGEROUS HOMs \\ \hline\hline
ELETTRA\hspace{1pt}2D & 500.461 & 5.65 & 3.56 & 79.9 & \multicolumn{1}{r|}{1.4205\hspace{2pt}GHz,\hspace{3pt}1.8768\hspace{2pt}GHz,\hspace{3pt}2.1257\hspace{2pt}GHz} \\ \hline
CAVITY~\#1 & \cellcolor{gray!30!white}\hspace{-3pt} 499.654 & \cellcolor{gray!30!white}\hspace{-3pt} 5.00 & \cellcolor{gray!30!white}\hspace{-3pt} 3.88 & 81.3 & - \\ \hline
CAVITY~\#2 & \cellcolor{gray!30!white}\hspace{-3pt} 499.654 & \cellcolor{gray!30!white}\hspace{-3pt} 4.98 & \cellcolor{gray!30!white}\hspace{-3pt} 3.83 & 81.0 & \cellcolor{gray!30!white}\hspace{-3pt} - \\ \hline
CAVITY~\#3 & \cellcolor{gray!30!white}\hspace{-3pt} 499.654 & \cellcolor{gray!30!white}\hspace{-3pt} 4.84 & \cellcolor{gray!30!white}\hspace{-3pt} 3.63 & 77.2 & \cellcolor{gray!30!white}\hspace{-3pt} - \\ \hline
\end{tabular}
\end{table}


The first multi-objective optimization problem we solve is
\begin{align}
& \min F_i(\boldsymbol{d}),\hspace{8pt} i=1,\dots,3,
\label{eq:moop-1}
\end{align}
where the objective functions are given below.
\begin{itemize}
\item[1.] $F_1(\boldsymbol{d}) = \big|f^{(0)}(\boldsymbol{d}) - f_{RF}\big|$, where $f_{RF} = 499.654\text{~MHz.}$
\item[2.] We maximize the shunt impedance~\cite{wiedemann2015particle} of the accelerating mode, i.e., we set
$$F_2(\boldsymbol{d}) = - R_{sh}^{(0)}(\boldsymbol{d}) = - \frac{\Big(V_{acc}^{(0)}(\boldsymbol{d})\Big)^2}{2\cdot P_{loss}(\boldsymbol{d})}.$$
The accelerating voltage $V_{acc}^{(0)}(\boldsymbol{d})$ is given by
\begin{align*}
& V_{acc}^{(0)}(\boldsymbol{d}) = \Bigg| \displaystyle\int_{0}^{L} E_z(0,z) e^{\;j\sqrt{\lambda}z}dz\Bigg|,
\end{align*}
where $L$ is the length of the cavity (see figure~\ref{fig:sym-elliptical-mesh}), $j$ the imaginary unit, $\lambda = \lambda^{(0)}(\boldsymbol{d})$ the computed eigenvalue, and $E_z$ the component of the electric field of the accelerating mode $E^{(0)}(\boldsymbol{d})$ in the direction of the axis of rotation (the orientation of the axes is shown in figure~\ref{fig:parameterization}). The power loss $P_{loss}(\boldsymbol{d})$ is defined as
\begin{align*}
& P_{loss}(\boldsymbol{d}) = \frac{R_{S}(\boldsymbol{d})}{2} \displaystyle\int_{\partial{\Omega}}|H_t|^2 dS,
\end{align*}
where $R_{S}(\boldsymbol{d})$ is the surface resistance, $\Omega = \Omega(\boldsymbol{d})$ the 3D domain, and $H_t$ the tangential component of the magnetic field of the accelerating mode $H^{(0)}(\boldsymbol{d})$.
We use $58.58\cdot 10^{6}$~S/m for the electrical conductivity of Cu-OFE~\cite{copper-conductivity}.
\item[3.] We minimize the normalized peak electric field on the cavity surface for the accelerating mode i.e., with $E = E^{(0)}(\boldsymbol{d})$,
\begin{align*}
F_3(\boldsymbol{d}) = \frac{\displaystyle\max_{\boldsymbol{x}\in\partial{\Omega}} \big|E(\boldsymbol{x})\big|}{V_{acc}^{(0)}(\boldsymbol{d})}.
\end{align*}
\end{itemize}
Since the cross section is symmetric and we are only interested in the properties of the accelerating mode, it is sufficient to use only half of the cross section and consider only the case with $m = 0$ and the PEC boundary conditions on the symmetry plane.
The number of triangles in the mesh is around 200'000, and, in order to ensure that the accelerating mode is found, the number of computed eigenpairs is $k_{max}=3$.

\begin{figure}
\begin{center}
\begin{minipage}[b]{\textwidth}
\includegraphics[width=\textwidth,trim={4cm 3.5cm 4.3cm 4.5cm}]{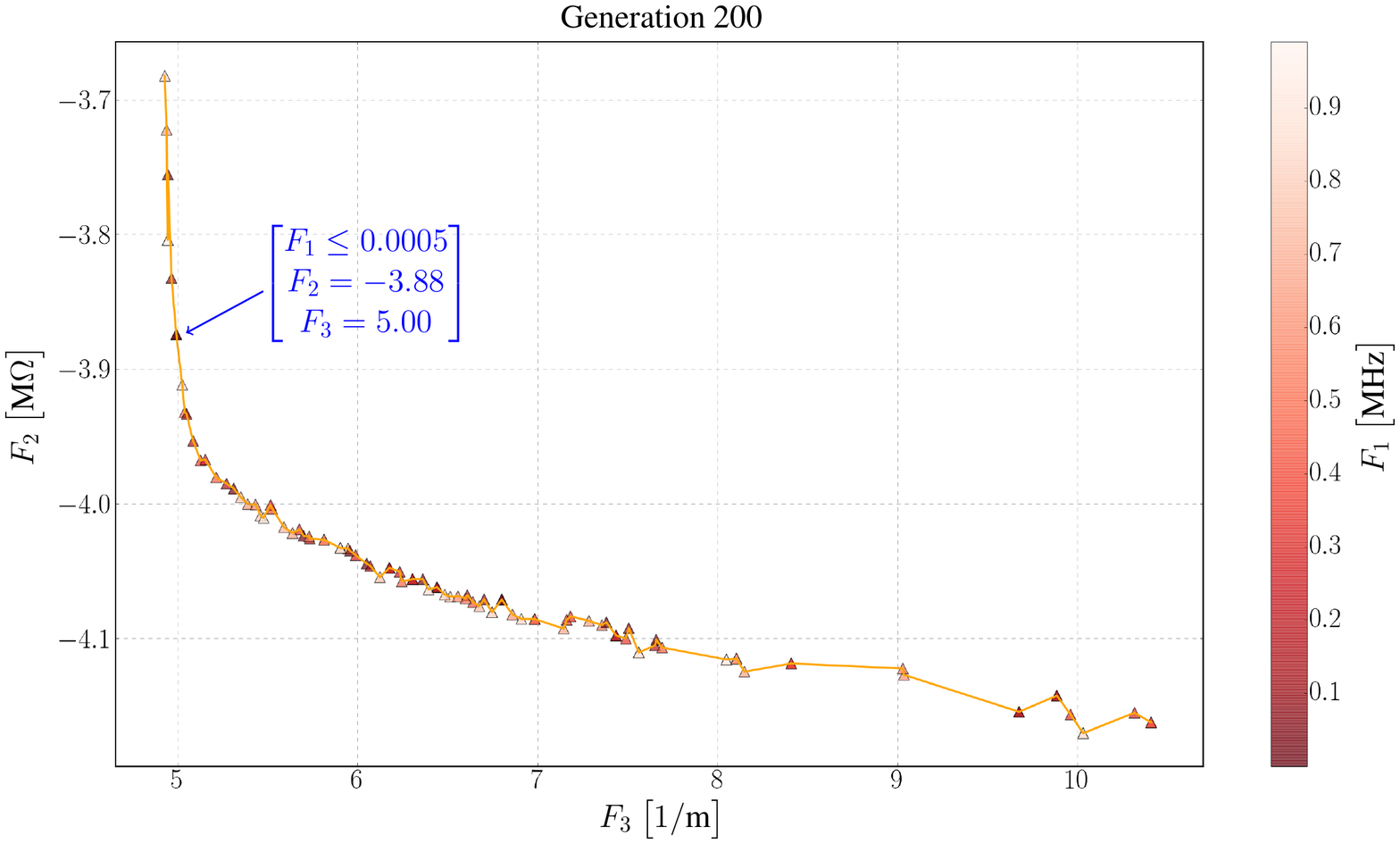}
\caption{The 200-th generation in the optimization of $F_1$, $F_2$, and $F_3$ using the geometry type from figure \ref{fig:sym-elliptical-mesh}. The arrow points to the individual whose accelerating electric field is shown in figure~\ref{fig:geom-type-1-id}.}
\label{fig:geom-type-1-gen}
\end{minipage}

\begin{minipage}[b]{\textwidth}
\includegraphics[width=\textwidth,trim={4cm 4.6cm 3.5cm 4cm}]{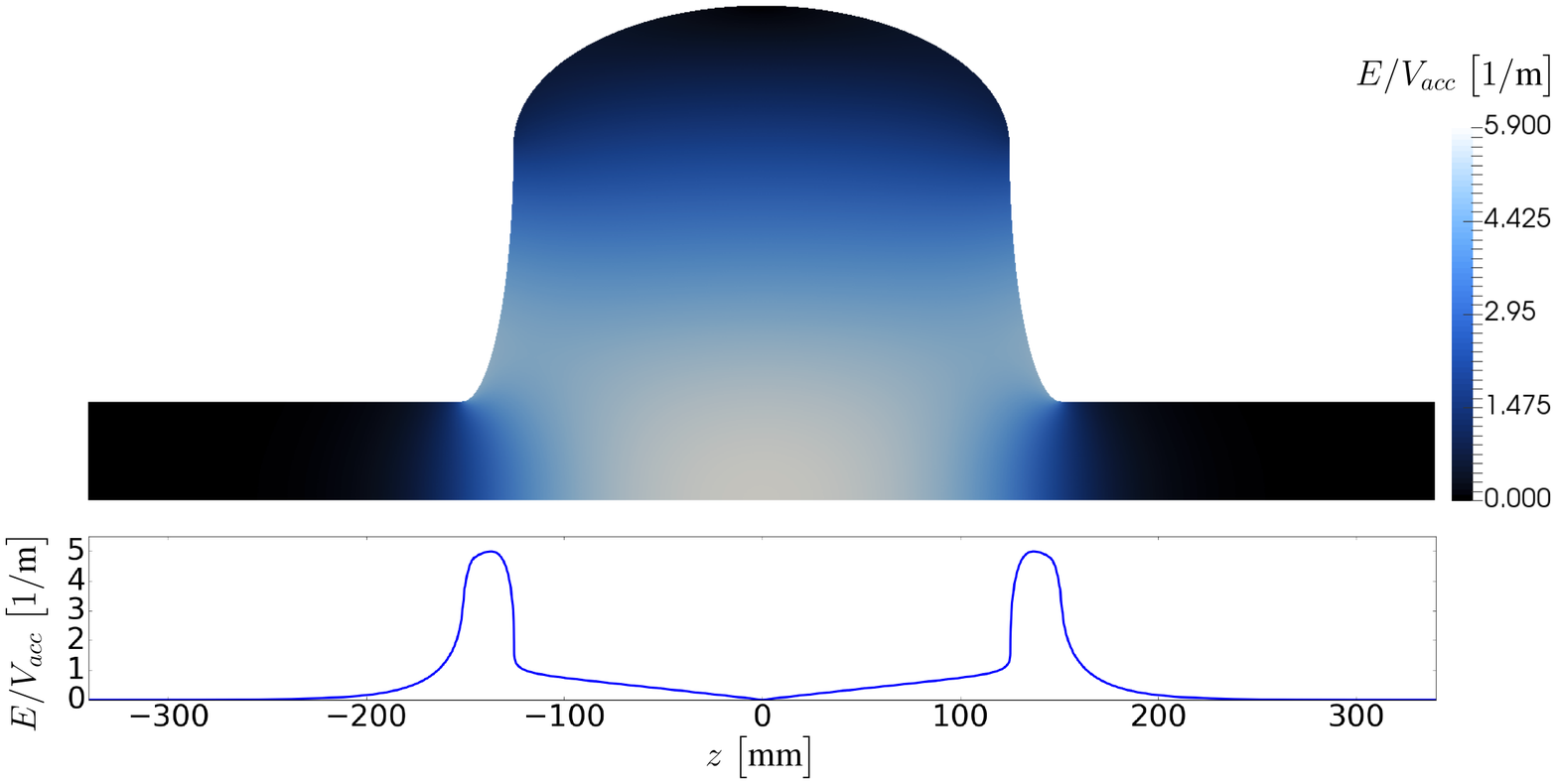}
\caption{Top: the magnitude of the axisymmetric electric field $E/V_{acc}$ for the accelerating mode of the cavity corresponding to the individual pointed at in figure~\ref{fig:geom-type-1-gen}, called CAVITY~\#1 in tables~\ref{table:individuals-description} and \ref{table:individuals-comparison}. Bottom: The magnitude of $E/V_{acc}$ on the cavity surface.}
\label{fig:geom-type-1-id}
\end{minipage}
\end{center}
\end{figure}

Solving the multi-objective optimization problem~(\ref{eq:moop-1}) rarely results in a cavity with the desired accelerating frequency, so we also formulate the requirement that $f^{(0)}$ should match $f_{RF}$ as the constraint~(\ref{eq:soft-frequency-constraint}) and solve the constrained multi-objective optimization problem
\begin{equation*}
\begin{aligned}
& \min F_i(\boldsymbol{d}), \hspace{8pt} i=1,\dots,3,\\
& H(\boldsymbol{d}) \leq \varepsilon,
\label{eq:moop-2}
\end{aligned}
\end{equation*}
with $\varepsilon = 1~$MHz.
The 200-th generation is shown in figure~\ref{fig:geom-type-1-gen}. 
Each triangle represents an individual in the generation, i.e., an RF cavity shape, and the orange line the computed Pareto front approximation.
The $x$ and $y$ coordinates represent the values of the objective functions $F_3$ and $F_2$, respectively, and the color shows the value of $F_1$. 
For all individuals, the accelerating frequency differs from the target frequency by less than 1~MHz, and the inverse correlation of $F_2$ and $F_3$ can be observed.
The marked individual possesses good objective function values. Its description is given in table~\ref{table:individuals-description}, where it is assigned the name CAVITY~\#1. The table contains the values of its design variables, as well as references to figures which contain additional information. The objective function values are listed in table~\ref{table:individuals-comparison}, where the shaded field indicates that the value was an objective in the corresponding optimization. The accelerating frequency of CAVITY~\#1 matches the given target frequency $f_{RF}$, and the values of $F_2$ and $F_3$ are better than the corresponding  values for ELETTRA~2D, i.e., the shunt impedance is higher and the peak value of the electric field on the cavity surface lower. 
Since ELETTRA~2D is only an axisymmetric approximation of the cavity used at SLS, we avoid comparing the frequencies.
The shape of CAVITY~\#1, as well as the electric field of the accelerating mode are shown in figure~\ref{fig:geom-type-1-id}.
Using two Intel XeonE5\_2680v3 nodes (each with 2$\times$12 cores @ 2.5~GHz and cache size 30~MB), it takes 4 hours and 50 minutes to compute 200 generations.


We now add another objective function, corresponding to the excitation of coupled beam modes, i.e., we next solve 
\begin{equation*}
\begin{aligned}
& \min F_i(\boldsymbol{d}),\hspace{8pt} i=1,\dots,4,\\
& H(\boldsymbol{d}) \leq \varepsilon,
\label{eq:moop-3}
\end{aligned}
\end{equation*}
where the new objective function, $F_4$, is defined below.

\begin{itemize}
\item[4.] For a cavity whose geometry is defined by the design point $\boldsymbol{d}$, with $p\in\mathbb{N}$, $q\in\{0,1,\dots,483\}$, and $f = p f_{RF}\pm (qf_0 + f_S)$, where
\begin{align*}
f_0 & =  f_{RF}/484 && \text{is the rotation frequency,}\\
f_S & =  2.5~\text{kHz} && \text{the synchrotron frequency without the harmonic cavity,}
\end{align*}
and $i>0$ the index of the longitudinal HOM, the longitudinal growth rate is given by~\cite{Chao:1993zn}
\begin{align*}
\frac{1}{\tau_{l,f}^{(i)}(\boldsymbol{d})} = \frac{|\alpha|}{2E/e}\frac{f_0}{f_S}I_{b}f^{(i)}(\boldsymbol{d}) \frac{R_{sh}^{(i)}(\boldsymbol{d})}{1+\Big(2Q_{0}^{(i)}(\boldsymbol{d})\frac{f^{(i)}(\boldsymbol{d}) - f}{f^{(i)}(\boldsymbol{d})}\Big)^2},
\end{align*}
where\vspace{-1pt}
\begin{align*}
\alpha & = -1.333\cdot 10^{-4} && \text{is the momentum compaction factor,}\\
E/e & =  2.4\cdot 10^9~\text{V} && \text{the beam energy divided by the electron charge,}\\
I_{b} & =  400~\text{mA} && \text{the beam current,}
\end{align*}
and $f^{(i)}(\boldsymbol{d})$, $R_{sh}^{(i)}(\boldsymbol{d})$ and $Q_{0}^{(i)}(\boldsymbol{d})$ are the frequency, shunt impedance and quality factor of the $i$-th HOM of that cavity, respectively. Denoting by $f_c$ the cutoff frequency, we define 
\begin{align*}
F_4(\boldsymbol{d}) = \sum_{\big\{i : f^{(i)}(\boldsymbol{d}) < f_c\big\}}\sum_{\substack{\big\{f = p f_{RF}\pm (qf_0 + f_S): \\p\in\{0,\dots,p_{max}\},\\ q \in\{ 0,1,\dots,483\}\big\}}} a_{f}^{(i)}(\boldsymbol{d}),
\end{align*}
where $p_{max}$ depends on $f_c$,
\begin{align*}
a_{f}^{(i)}(\boldsymbol{d}) = 
\begin{cases}
\displaystyle\frac{1}{\tau_{l,f}^{(i)}(\boldsymbol{d})}, & \text{ if }\displaystyle\frac{1}{\tau_{l,f}^{(i)}(\boldsymbol{d})} \geq  b,\\
0, & \text{ otherwise},
\end{cases}
\end{align*} 
\begin{equation}
b = 0.5\cdot \frac{1}{\tau_E},
\label{eq:b}
\end{equation}
where $0.5$ is a safety factor and $\tau_E = 6.5~\text{ms}$ is the longitudinal damping time for SLS-2.
With $r = 50$~mm, $f_{c}=2.29~$GHz and $p_{max} = 5$.
\end{itemize}

\begin{figure}
\begin{center}
\begin{minipage}[b]{\textwidth}
\includegraphics[width=\textwidth,trim={4cm 3.5cm 4.3cm 4.5cm}]{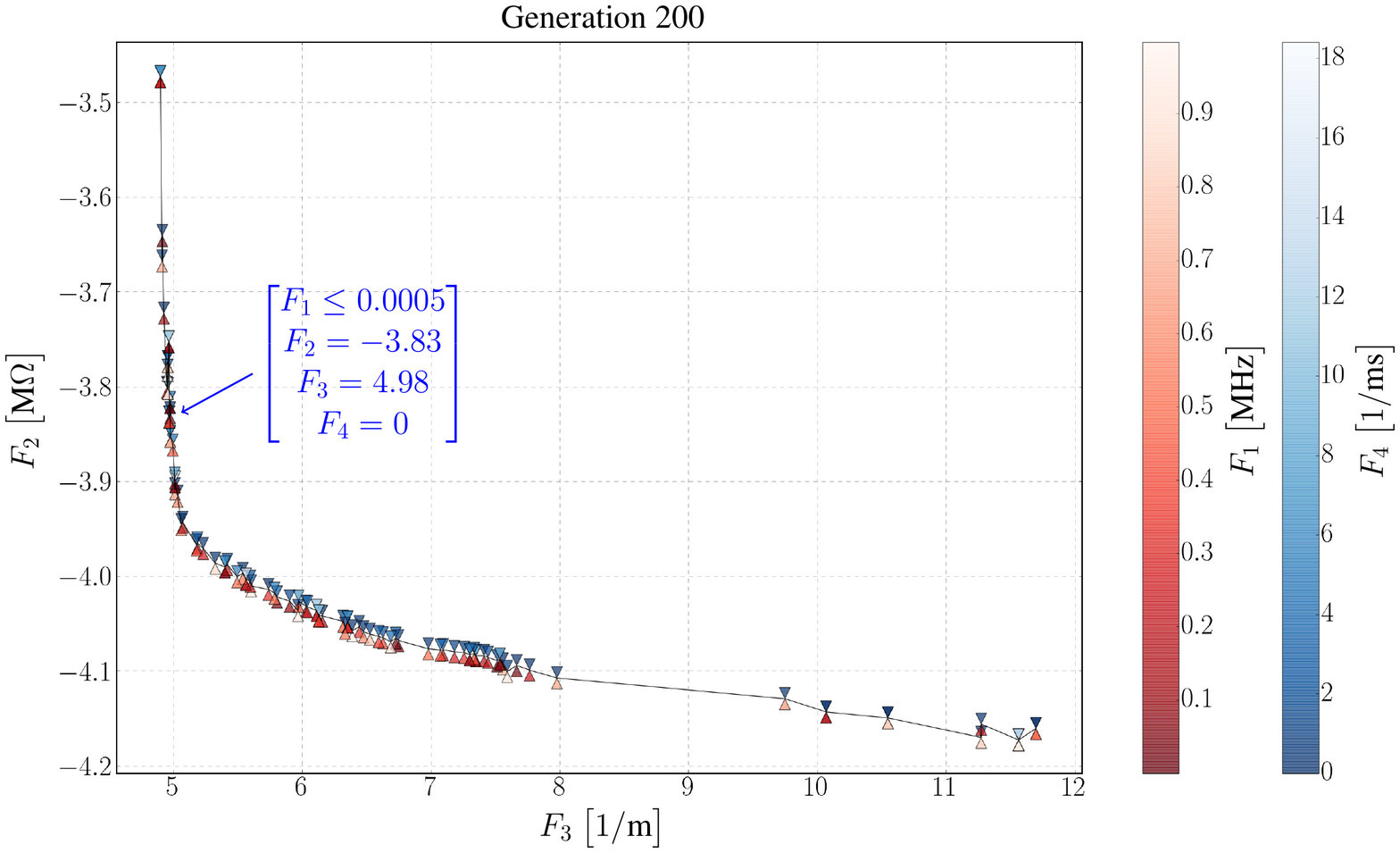}
\caption{The 200-th generation in the optimization of $F_1$, $F_2$, $F_3$ and $F_4$ using the geometry type from figure \ref{fig:sym-elliptical-mesh}. The arrow points to the individual whose accelerating electric field is shown in figure~\ref{fig:geom-type-1-id-4obj}.}
\label{fig:geom-type-1-gen-4obj}
\end{minipage}

\begin{minipage}[b]{\textwidth}
\includegraphics[width=\textwidth,trim={4cm 4.6cm 3.5cm 4cm}]{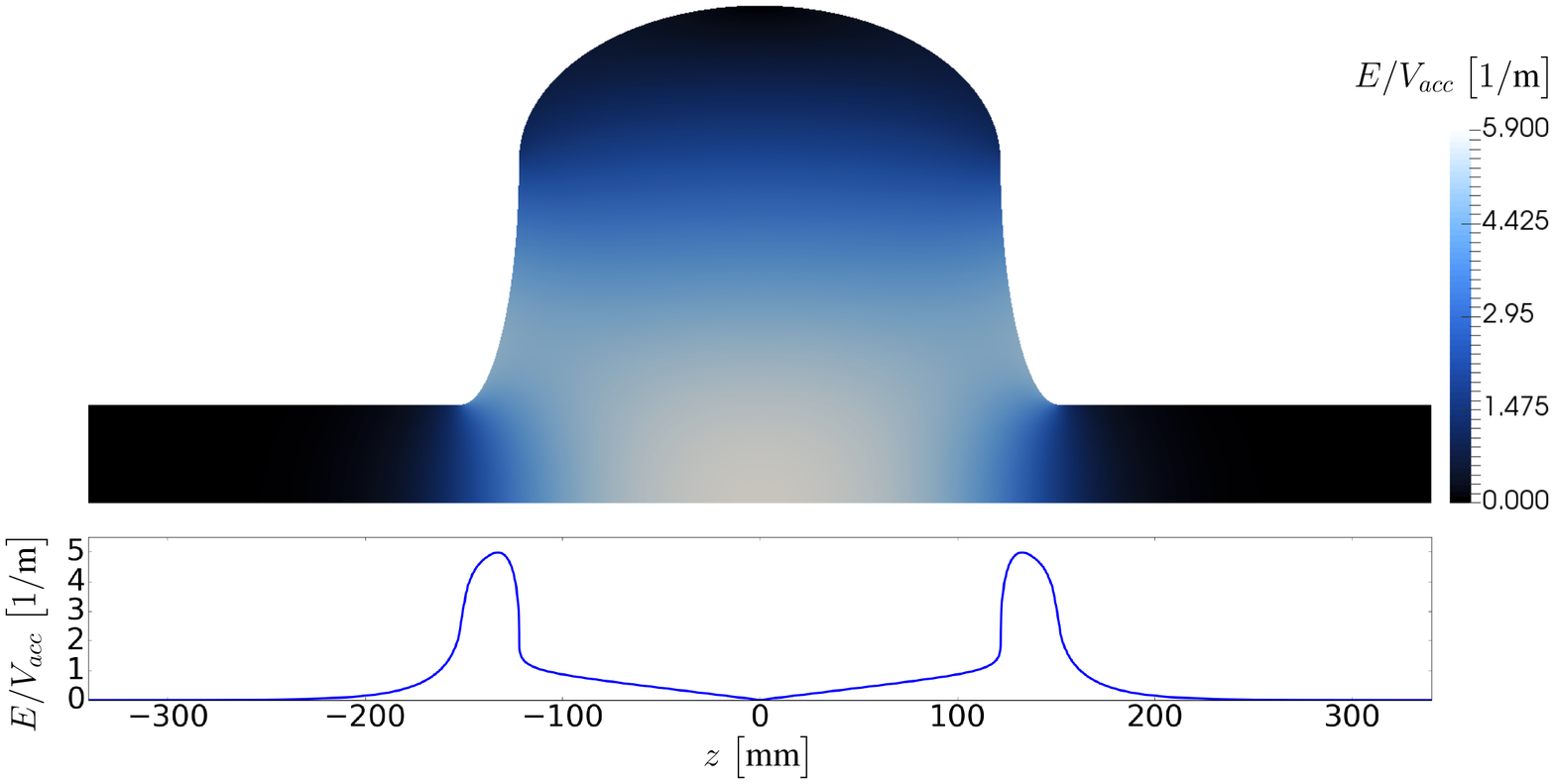}
\caption{Top: the magnitude of the axisymmetric electric field $E/V_{acc}$ for the accelerating mode of the cavity corresponding to the individual pointed at in figure~\ref{fig:geom-type-1-gen-4obj}, called CAVITY~\#2 in tables~\ref{table:individuals-description} and \ref{table:individuals-comparison}. Bottom: The magnitude of $E/V_{acc}$ on the cavity surface.}
\label{fig:geom-type-1-id-4obj}
\end{minipage}
\end{center} 
\end{figure}

We consider only longitudinal modes, i.e., only the monopole case $m = 0$. Since we are again using only half of the cross section, it is necessary to consider both the case with the PEC and the case with the PMC boundary conditions. 
In order to achieve the necessary accuracy for $F_4$, the number of triangles in the mesh is now around 400'000, and, in order for all the longitudinal modes with frequencies below $f_{c}$ to be computed, $k_{max}=20$.
In this case it takes 2 days and 18 hours to compute 200 generations on two nodes, since the mesh size and the number of computed eigenpairs are larger, and we need to solve twice as many eigenproblems. Using eight nodes, however, brings the time down to 18 hours and 20 minutes, which amounts to the speedup of 3.6.

The 200-th generation is shown in figure~\ref{fig:geom-type-1-gen-4obj}. The red and blue triangles show the values of $F_1$ and $F_4$, respectively, and the $x$ and $y$ coordinates of the point where they meet represent the values of the objective functions $F_3$ and $F_2$, respectively. 
For all individuals, the accelerating frequency again differs from the target frequency by less than 1~MHz, and the inverse correlation of $F_2$ and $F_3$ can again be observed. An individual with good objective function values is chosen and marked, and its descriptions and objective function values are given in tables~\ref{table:individuals-description} and \ref{table:individuals-comparison}, respectively, where it is referred to as CAVITY~\#2. 
Its accelerating frequency matches the target frequency, the peak electric field on the cavity surface is slightly lower than for CAVITY~\#1, i.e., $F_3$ is slightly better, but the shunt impedance is not as high, i.e., $F_2$ is not as good, though it is still better than the corresponding value for ELETTRA~2D.
Since ELETTRA~2D is only an axisymmetric approximation, the longitudinal growth rate is above the threshold $b$, defined in (\ref{eq:b}), for three of its HOMs. 
The value of $F_4$ is zero for CAVITY~\#2, and, coincidentally, also for CAVITY~\#1, even though it was not an objective in the optimization.\\

\begin{figure}[h!]
\includegraphics[width=\textwidth,trim={4cm 4.9cm 3.5cm 4.7cm}]{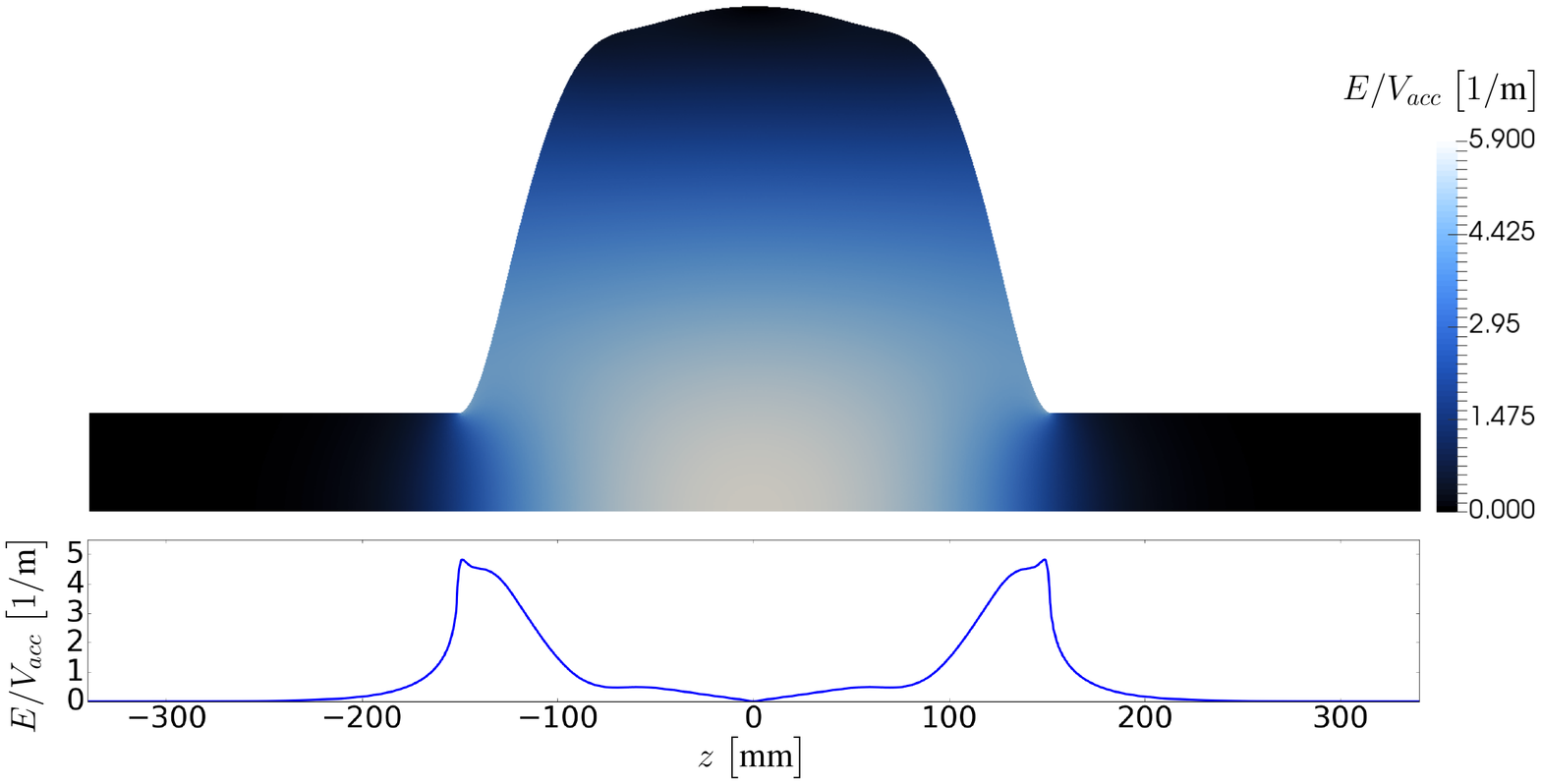}
\caption{Top: the magnitude of the axisymmetric electric field $E/V_{acc}$ for the accelerating mode of the cavity referred to as CAVITY~\#3 in tables~\ref{table:individuals-description} and \ref{table:individuals-comparison}. Bottom: The magnitude of $E/V_{acc}$ on the cavity surface.}
\label{fig:geom-type-3-id2-4obj}
\end{figure}

We now consider the geometry type illustrated in figure~\ref{fig:splines}. The symmetric cross section is defined as a complete cubic spline with equidistant knots and horizontal end slopes. We fix the number of knots to 6, and fix the values of $L$, $l$, and $r = y_0$. This leaves us with five design variables, $y_1,\dots,y_5$. 
Choosing the initial design variables in the ranges $y_1\in[100,200]$, $y_2\in[150,250]$, $y_3\in[200,250]$, $y_4\in[250,275]$, $y_5\in[250,275]$, we find, for example, CAVITY~\#3, which is shown in figure~\ref{fig:geom-type-3-id2-4obj}, described in table~\ref{table:individuals-description}, and whose objective function values are given in table~\ref{table:individuals-comparison}.
Its accelerating frequency matches the target frequency, $F_4 = 0$, the peak electric field on the cavity surface is slightly lower than for CAVITY~\#1 and CAVITY~\#2, i.e., $F_3$ is slightly better, but the shunt impedance is not as high, i.e., $F_2$ is not as good, though it is still slightly better than the corresponding value for ELETTRA~2D.

\section{Conclusions and future work}\label{sec:conclusions}
In this paper we tackled the problem of RF cavity shape optimization using a multi-objective evolutionary algorithm. 
As an example problem we solved the hypothetical problem of optimizing the shape of the main RF cavity of the proposed SLS upgrade, SLS-2. 
Under the assumption of axisymmetry, we found interesting new RF cavity shapes with the desired accelerating frequency that better meet all of the objectives than the axisymmetric approximation of the current cavity of SLS.
We considered different objectives and different geometry types, and showed the benefit of employing a penalty method for dealing with the constraint on the accelerating frequency.
Other geometry types and objectives can easily be added, and, at the expense of increased computational cost, the same optimization method could be employed with a fully three-dimensional solver. 
In the future, the benefit of using other constraint handling techniques, gradient information, or some form of local search could also be explored.


\section*{Acknowledgements}

We executed the computations on the Euler compute cluster of ETH~Zurich, at the expense of a PSI grant.




\begin{thebibliography}{10}
\expandafter\ifx\csname url\endcsname\relax
  \def\url#1{\texttt{#1}}\fi
\expandafter\ifx\csname urlprefix\endcsname\relax\def\urlprefix{URL }\fi
\expandafter\ifx\csname href\endcsname\relax
  \def\href#1#2{#2} \def\path#1{#1}\fi

\bibitem{ising1924prinzip}
G.~Ising, {Prinzip einer Methode zur Herstellung von Kanalstrahlen hoher
  Voltzahl}, Ark. Mat. Astr. Fys., Almqvist \& Wiksells Boktryckeri AB, 1924.

\bibitem{Wideroe:1928flt}
R.~Wider\"oe, {\"Uber ein neues Prinzip zur Herstellung hoher Spannungen},
  Arch. Elektrotech. 21~(4) (1928) 387--406.
\newblock \href {http://dx.doi.org/10.1007/BF01656341}
  {\path{doi:10.1007/BF01656341}}.

\bibitem{PhysRev.40.19}
E.~O. Lawrence, M.~S. Livingston, The production of high speed light ions
  without the use of high voltages, Phys. Rev. 40 (1932) 19--35.
\newblock \href {http://dx.doi.org/10.1103/PhysRev.40.19}
  {\path{doi:10.1103/PhysRev.40.19}}.

\bibitem{Yamada:2008zzb}
K.~Yamada, et~al., {Status of the Superconducting Ring Cyclotron at RIKEN RI
  Beam Factory}, in: {Proc. 11th European Particle Accelerator Conference
  (EPAC2008)}, Vol. C0806233, 2008, p. THPP069.

\bibitem{Fernandes:1989dn}
P.~Fernandes, R.~Massarino, R.~Parodi, A.~Tarditi, A.~Massarotti, {The design
  of the RF cavities for Elettra}, in: Proc. 1989 IEEE Particle Accelerator
  Conference, Vol.~1, 1989, pp. 220--222.

\bibitem{Svandrlik:2000gf}
M.~Svandrlik, et~al., {The SUPER-3HC project: An idle superconducting harmonic
  cavity for bunch length manipulation}, in: {Proc. 7th European Particle
  Accelerator Conference (EPAC2000)}, Vol. 1-3, 2000, pp. 2052--2054.

\bibitem{PhysRevAccelBeams.19.100702}
T.~Schietinger, et~al., {Commissioning experience and beam physics measurements
  at the SwissFEL Injector Test Facility}, Phys. Rev. Accel. Beams 19 (2016)
  100702.
\newblock \href {http://dx.doi.org/10.1103/PhysRevAccelBeams.19.100702}
  {\path{doi:10.1103/PhysRevAccelBeams.19.100702}}.

\bibitem{Sun:2015eta}
J.~Sun, P.-A. Duperrex, G.~Kotrle, Design of a new beam current monitor under
  heavy heat load, in: {Proc. 54th ICFA Advanced Beam Dynamics Workshop on
  High-Intensity and High-Brightness Hadron Beams (HB2014)}, 2015, p. MOPAB48.

\bibitem{Keil:2016pqq}
B.~Keil, et~al., {Status of the SwissFEL BPM System}, in: {Proc. 4th
  International Beam Instrumentation Conference (IBIC2015)}, 2016, p. TUPB065.
\newblock \href {http://dx.doi.org/10.18429/JACoW-IBIC2015-TUPB065}
  {\path{doi:10.18429/JACoW-IBIC2015-TUPB065}}.

\bibitem{Marchetti:2017eym}
B.~Marchetti, et~al., {X-Band TDS Project}, in: {Proc. 8th International
  Particle Accelerator Conference (IPAC2017)}, 2017, p. MOPAB044.
\newblock \href {http://dx.doi.org/10.18429/JACoW-IPAC2017-MOPAB044}
  {\path{doi:10.18429/JACoW-IPAC2017-MOPAB044}}.

\bibitem{SLS2CDR2017}
A.~Streun,
  \href{http://www.lib4ri.ch/archive/nebis/PSI_Berichte_000478272/PSI-Bericht_17-03.pdf}{{SLS-2}
  conceptual design report}, PSI Bericht 17-03, PSI, 5232 Villigen, Switzerland
  (December 2017).

\bibitem{wiedemann2015particle}
H.~Wiedemann, Particle Accelerator Physics, Graduate Texts in Physics,
  Springer, Berlin, 2015.

\bibitem{jackson_classical_1999}
J.~D. Jackson, Classical Electrodynamics, 3rd Edition, Wiley, New York, 1999.

\bibitem{Halbach:1976cp}
K.~Halbach, R.~F. Holsinger, Superfish -- a computer program for evaluation of
  {RF} cavities with cylindrical symmetry, Part. Accel. 7 (1976) 213--222.

\bibitem{GeusPhd}
R.~Geus, The {J}acobi--{D}avidson algorithm for solving large sparse symmetric
  eigenvalue problems with application to the design of accelerator cavities,
  Ph.D. thesis, ETH Z\"{u}rich (Diss. ETH No. 14734) (2002).

\bibitem{hfss}
{ANSYS}\textregistered\ {HFSS},
  \url{https://www.ansys.com/Products/Electronics/ANSYS-HFSS}.

\bibitem{MicrowaveStudio}
{CST Microwave Studio}\textregistered,
  \url{https://www.cst.com/products/cstmws}.

\bibitem{ansys}
{ANSYS}\textregistered\ {Multiphysics},
  \url{https://www.ansys.com/products/platform/multiphysics-simulation}.

\bibitem{comsol}
{COMSOL Multiphysics}\textregistered,
  \url{https://www.comsol.com/comsol-multiphysics}.

\bibitem{InjectorIIResonatorCC07}
L.~Stingelin, M.~Bopp, H.~Fitze, {Development of the new 50MHz resonators for
  the PSI Injector II cyclotron}, in: Proc. 18th International conference on
  cyclotrons and their applications, 2007, pp. 467--469.

\bibitem{ChinellatoPhd}
O.~Chinellato, The complex-symmetric {J}acobi--{D}avidson algorithm and its
  application to the computation of some resonance frequencies of anisotropic
  lossy axisymmetric cavities, Ph.D. thesis, ETH Z\"{u}rich (Diss. ETH No.
  16243) (2005).

\bibitem{arbenz:01}
P.~Arbenz, R.~Geus, S.~Adam, Solving {M}axwell eigenvalue problems for
  accelerating cavities, Phys. Rev. ST Accel. Beams 4~(2) (2001) 022001.
\newblock \href {http://dx.doi.org/10.1103/PhysRevSTAB.4.022001}
  {\path{doi:10.1103/PhysRevSTAB.4.022001}}.

\bibitem{Schaer:2016jhu}
M.~Schaer, et~al., {RF traveling-wave electron gun for photoinjectors}, Phys.
  Rev. Accel. Beams 19~(7) (2016) 072001.
\newblock \href {http://dx.doi.org/10.1103/PhysRevAccelBeams.19.072001}
  {\path{doi:10.1103/PhysRevAccelBeams.19.072001}}.

\bibitem{akcelik:05}
V.~{Ak{\c c}elik}, et~al., Adjoint methods for electromagnetic shape
  optimization of the low-loss cavity for the {International Linear Collider},
  J. Phys. Conf. Ser. 16~(1) (2005) 435.

\bibitem{akcelik:08}
V.~{Ak{\c c}elik}, et~al., Shape determination for deformed electromagnetic
  cavities, J. Comput. Phys 227~(3) (2008) 1722 -- 1738.
\newblock \href {http://dx.doi.org/10.1016/j.jcp.2007.09.029}
  {\path{doi:10.1016/j.jcp.2007.09.029}}.

\bibitem{akcelik:09}
V.~{Ak{\c c}elik}, et~al., Large scale shape optimization for accelerator
  cavities, J. Phys. Conf. Ser. 180~(1) (2009) 012001.

\bibitem{kirkpatric:83}
S.~Kirkpatrick, D.~Gelatt~Jr, M.~P.~Vecchi, Optimization by simulated
  annealing, Science 220 (1983) 671--680.

\bibitem{kennedy:95}
J.~Kennedy, R.~Eberhart, Particle swarm optimization, in: Proc. IEEE
  International Conference on Neural Networks (ICNN1995), Vol.~4, 1995, pp.
  1942--1948.

\bibitem{kalyanmoy-deb}
K.~Deb, Multi-Objective Optimization Using Evolutionary Algorithms, Wiley, New
  York, 2001.

\bibitem{dan-simon}
D.~Simon, Evolutionary Optimization Algorithms, Wiley, New York, 2013.

\bibitem{bazarov:05}
I.~V. Bazarov, C.~K. Sinclair, Multivariate optimization of a high brightness
  dc gun photoinjector, Phys. Rev. ST Accel. Beams 8~(3) (2005) 034202.
\newblock \href {http://dx.doi.org/10.1103/PhysRevSTAB.8.034202}
  {\path{doi:10.1103/PhysRevSTAB.8.034202}}.

\bibitem{ineichen:13}
Y.~Ineichen, A.~Adelmann, C.~Bekas, A.~Curioni, P.~Arbenz, A fast and scalable
  low dimensional solver for charged particle dynamics in large particle
  accelerators, Comput. Sci. Res. Dev. 28~(2) (2013) 185--192.
\newblock \href {http://dx.doi.org/10.1007/s00450-012-0216-2}
  {\path{doi:10.1007/s00450-012-0216-2}}.

\bibitem{RevModPhys.18.441}
J.~C. Slater, Microwave electronics, Rev. Mod. Phys. 18 (1946) 441--512.
\newblock \href {http://dx.doi.org/10.1103/RevModPhys.18.441}
  {\path{doi:10.1103/RevModPhys.18.441}}.

\bibitem{francavilla}
A.~Francavilla, C.~V.~Ramakrishnan, O.~C~Zienkiewicz, Optimization of shape to
  minimize stress concentration, J. Strain Anal. Eng. Des. 10 (1975) 63--70.

\bibitem{le}
C.~Le, T.~Bruns, D.~Tortorelli, A gradient-based, parameter-free approach to
  shape optimization, Comput. Methods Appl. Mech. Eng. 200 (2011) 985--996.

\bibitem{bhavikatti}
S.~S.~Bhavikatti, C.~V.~Ramakrishnan, Optimum shape design of rotating disks,
  Comput. Struct. 11 (1980) 397--401.

\bibitem{tortorelli-2}
D.~A. Tortorelli, J.~A. Tomasko, T.~E. Morthland, J.~A. Dantzig, Optimal design
  of nonlinear parabolic systems. {Part II}: Variable spatial domain with
  applications to casting optimization, Comput. Methods Appl. Mech. Eng. 113
  (1994) 157--172.

\bibitem{braibant}
V.~Braibant, C.~Fleury, Shape optimal design using {B}-splines, Comput. Methods
  Appl. Mech. Eng. 44 (1984) 247--267.

\bibitem{haslinger}
J.~Haslinger, R.~A.~E. Makinen, Introduction to Shape Optimization: Theory,
  Approximation, and Computation, SIAM, Philadelphia, 2003.

\bibitem{gmsh}
C.~Geuzaine, J.-F. Remacle, Gmsh: a three-dimensional finite element mesh
  generator with built-in pre- and post-processing facilities, Int. J. Numer.
  Methods Eng. 79~(11) (2009) 1309--1331.

\bibitem{jdsym-trilinos}
P.~Arbenz, M.~Be\v{c}ka, R.~Geus, U.~Hetmaniuk, T.~Mengotti, On a parallel
  multilevel preconditioned {M}axwell eigensolver, Parallel Comput. 32~(2)
  (2006) 157--165.

\bibitem{abgh:04}
P.~Arbenz, M.~Be\v{c}ka, R.~Geus, U.~Hetmaniuk, Towards a parallel multilevel
  preconditioned {M}axwell eigensolver, in: Proc. 7th International Conference
  on Applied Parallel Computing: State of the Art in Scientific Computing,
  Springer, Berlin, 2006, pp. 831--838, (Lect. Notes Comput. Sci., 3732).

\bibitem{trilinos}
M.~A. Heroux, et~al., An overview of the {Trilinos} project, ACM Trans. Math.
  Softw. 31~(3) (2005) 397--423.
\newblock \href {http://dx.doi.org/10.1145/1089014.1089021}
  {\path{doi:10.1145/1089014.1089021}}.

\bibitem{jin:93}
J.~Jin, The Finite Element Method in Electromagnetics, Wiley, New York, 1993.

\bibitem{nsga-ii}
K.~Deb, A.~Pratap, S.~Agarwal, T.~Meyarivan, A fast and elitist multiobjective
  genetic algorithm: {NSGA-II}, Trans. Evol. Comp 6~(2) (2002) 182--197.
\newblock \href {http://dx.doi.org/10.1109/4235.996017}
  {\path{doi:10.1109/4235.996017}}.

\bibitem{pisa}
S.~Bleuler, M.~Laumanns, L.~Thiele, E.~Zitzler, {PISA}: A platform and
  programming language independent interface for search algorithms, in: Proc.
  2nd International Conference on Evolutionary Multi-criterion Optimization
  (EMO2003), Springer, Berlin, 2003, pp. 494--508.

\bibitem{IneichenPhd}
Y.~Ineichen, Towards massively parallel multi-objective optimization with
  application to particle accelerators, Ph.D. thesis, ETH Z\"{u}rich (Diss. ETH
  No. 21114) (2013).

\bibitem{copper-conductivity}
ISO 431-1981 (E).

\bibitem{Chao:1993zn}
A.~W. Chao, {Physics of Collective Beam Instabilities in High-Energy
  Accelerators}, Wiley, New York, 1993.

\end{thebibliography}
\end{document}